\documentstyle[12pt]{article}

\begin{document}

\title{Gravitation, Thermodynamics, and Quantum Theory}
\author{Robert M. Wald\\
         {\it Enrico Fermi Institute and Department of Physics}\\
         {\it University of Chicago}\\
         {\it 5640 S. Ellis Avenue}\\
         {\it Chicago, Illinois 60637-1433}}
\maketitle

\begin{abstract}

During the past 30 years, research in general relativity has brought
to light strong hints of a very deep and fundamental relationship
between gravitation, thermodynamics, and quantum theory. The most
striking indication of such a relationship comes from black hole
thermodynamics, where it appears that certain laws of black hole
mechanics are, in fact, simply the ordinary laws of thermodynamics
applied to a system containing a black hole. This article will review
the present status of black hole thermodynamics and will discuss some
of the related unresolved issues concerning gravitation,
thermodynamics, and quantum theory.

\end{abstract}
\newpage

\section{Introduction}

At the end of a century---particularly one marking the end of a
millennium---it is natural to attempt to take stock of the status of a
field of endeavor from as broad a perspective as possible. In the
field of physics, two theories---general relativity and quantum
mechanics---were developed during the first quarter of the present
century. These theories revolutionized the way we think about the
physical world. Despite enormous progress during the remainder of this
century in exploring their consequences and in the application of
these theories to construct successful ``standard models'' of
cosmology and particle physics, at the end of this century we are
still struggling to reconcile general relativity and quantum theory at
a fundamental level.

The revolutions in physics introduced by general relativity and
quantum theory were accompanied by major changes in the language and
concepts used to describe the physical world. In general relativity,
it is recognized that space and time meld into a single
entity---spacetime---and that the structure of spacetime is described
by a Lorentz metric, which has a dynamical character. Consequently,
simple Newtonian statements such as ``two particles are a distance $d$
apart at time $t$'' become essentially meaningless in general
relativity until one states exactly how $t$ and $d$ are defined for
the particular class of spacetime metrics under
consideration. Furthermore, concepts such as the ``local energy
density of the gravitational field'' are absent in general
relativity. The situation is considerably more radical in quantum
theory, where the existence of ``objective reality'' itself is denied,
i.e., observables cannot, in general, consistently be assigned
definite values.

I believe that the proper description of quantum phenomena in strong
gravitational fields will necessitate revolutionary conceptual changes
in our view of the physical world at least comparable to those that
occurred in the separate developments of general relativity and
quantum theory. At present, the greatest insights into the
physical nature of quantum phenomena in strong gravitational fields
comes from the analysis of thermodynamic properties associated with
black holes.  This analysis also provides strong hints that
statistical physics may be deeply involved in any fundamental
conceptual changes that accompany a proper description of quantum
gravitational phenomena.

At the present time, string theory is the most promising candidate for
a theory of quantum gravity. One of the greatest successes of string
theory to date has been the calculation of the entropy of certain
classes of black holes. However, the formulation of string theory is
geared much more towards the calculation of scattering matrices in
asymptotically flat spacetimes rather than towards providing a local
description of physical phenomena in strong gravitational
fields. Within the framework of string theory, it is difficult to
imagine even how to pose (no less, how to calculate the answer to)
local physical questions like, ``What would an observer who falls into
a black hole experience as he approaches what corresponds classically
to the spacetime singularity within the black hole?''  Thus, string
theory has not yet provided us with new conceptual insights into the
physical nature of phenomena occurring in strong gravitational fields
that are commensurate with some of its mathematical successes. It may
well be that---even assuming it is a correct theory of nature---string
theory will bear a relationship to the ``ultimate theory of
everything'' that is similar to the relationship between ``old quantum
theory'' and quantum theory. Therefore, I feel that it is very
encouraging that, at the present time, intensive efforts are underway
toward providing reformulations of string theory. However, to date,
the these efforts have mainly been concerned with obtaining a
formulation that would unify the (different looking) versions of
string theory, rather than achieving new conceptual foundations for
describing local quantum phenomena occurring in strong gravitational fields.

Thus, at present, most of the physical insights into quantum phenomena
occurring in strong gravitational fields arise from classical and
semiclassical analyses of black holes in general relativity. In this
article, I will review classical and quantum black hole thermodynamics
and then discuss some unresolved issues and puzzles, which may provide
some hints as to the new conceptual features that may be present in
the quantum description of strong gravitational fields. In the
discussion, I will not attempt to provide a balanced account of
research in this area, but rather will merely present my own views.

\section{Classical black hole mechanics}

Undoubtedly, one of the most remarkable and unexpected developments in
theoretical physics to have occurred during the latter portion of this
century was the discovery of a close relationship between certain laws
of black hole physics and the ordinary laws of thermodynamics. It was
first pointed out by Bekenstein \cite{b1} that the area nondecrease
theorem of classical general relativity \cite{h} is analogous to the
ordinary second law of thermodynamics, and he proposed that the area,
$A$, of a black hole (times a constant of order unity in Planck units)
should be interpreted as its physical entropy. A short time
later, Bardeen, Carter, and Hawking \cite{bch} extended the analogy
between black holes and thermodynamics considerably further by proving
black hole analogs of the zeroth and first laws of thermodynamics. In
this section, I will give a brief review of the laws of classical
black hole mechanics.

First, we review the definition of a black hole and some properties of
stationary black holes. In physical terms, a black hole is a region
where gravity is so strong that nothing can escape. In order to make
this notion precise, one must have in mind a region of spacetime to
which one can contemplate escaping. For an asymptotically flat
spacetime $(M, g_{ab})$ (representing an isolated system), the
asymptotic portion of the spacetime ``near infinity'' is such a
region. The {\em black hole} region, ${\cal B}$, of an asymptotically
flat spacetime, $(M, g_{ab})$, is defined as
\begin{equation}
{\cal B} \equiv M - I^-({\cal I}^+) ,
\label{bh}
\end{equation}
where ${\cal I}^+$ denotes future null infinity and $I^-$ denotes the
chronological past.  The {\em event horizon}, ${\cal H}$, of a black
hole is defined to be the boundary of ${\cal B}$. In particular,
${\cal H}$ is a null hypersurface. Note that the entire future history
of the spacetime must be known before the location of ${\cal H}$ can
be determined, i.e., ${\cal H}$ possesses no distinguished local
significance.

If an asymptotically flat spacetime $(M, g_{ab})$ contains a black
hole, ${\cal B}$, then ${\cal B}$ is said to be {\em stationary} if
there exists a one-parameter group of isometries on $(M, g_{ab})$
generated by a Killing field $t^a$ which is unit timelike at
infinity. The black hole is said to be {\em static} if it is
stationary and if, in addition, $t^a$ is hypersurface orthogonal. The
black hole is said to be {\em axisymmetric} if there exists a one
parameter group of isometries which correspond to rotations at
infinity. A stationary, axisymmetric black hole is said to possess the
``$t - \phi$ orthogonality property'' if the 2-planes spanned by $t^a$
and the rotational Killing field $\phi^a$ are orthogonal to a family
of 2-dimensional surfaces.

A null surface, ${\cal K}$, whose null generators coincide with the
orbits of a one-parameter group of isometries (so that there is a
Killing field $\xi^a$ normal to ${\cal K}$) is called a {\em Killing
horizon}. There are two independent results (usually referred to as
``rigidity theorems'') that show that in wide variety of cases of
interest, the event horizon, ${\cal H}$, of a stationary black hole
must be a Killing horizon. The first, due to Carter \cite{c}, states
that for a static black hole, the static Killing field $t^a$ must be
normal to the horizon, whereas for a stationary-axisymmetric black
hole with the $t - \phi$ orthogonality property there exists a Killing
field $\xi^a$ of the form
\begin{equation}
\xi^a = t^a + \Omega \phi^a
\label{xi}
\end{equation}
which is normal to the event horizon. The constant $\Omega$ defined by
eq.(\ref{xi}) is called the {\em angular velocity of the horizon}.
Carter's result does not rely on any field equations, but leaves open
the possibility that there could exist stationary black holes without
the above symmetries whose event horizons are not Killing
horizons. The second result, due to Hawking \cite{he} (see also
\cite{frw}), directly proves that in vacuum or electrovac general
relativity, the event horizon of any stationary black hole must be a
Killing horizon. Consequently, if $t^a$ fails to be normal
to the horizon, then there must exist an additional Killing field,
$\xi^a$, which is normal to the horizon, i.e., a stationary black hole
must be nonrotating (from which staticity follows \cite{sw},
\cite{cw}) or axisymmetric (though not necessarily with the $t - \phi$
orthogonality property). Note that Hawking's theorem makes no
assumptions of symmetries beyond stationarity, but it does rely on the
properties of the field equations of general relativity.

Now, let ${\cal K}$ be any Killing horizon (not necessarily required
to be the event horizon, ${\cal H}$, of a black hole), with normal
Killing field $\xi^a$. Since $\nabla^a (\xi^b \xi_b)$ also is normal
to ${\cal K}$, these vectors must be proportional at every point on
${\cal K}$. Hence, there exists a function, $\kappa$, on ${\cal K}$,
known as the {\em surface gravity} of ${\cal K}$, which is defined by
the equation
\begin{equation}
\nabla^a (\xi^b \xi_b) = -2 \kappa \xi^a
\label{kappa}
\end{equation}
It follows immediately that $\kappa$ must be constant along each null
geodesic generator of ${\cal K}$, but, in general, $\kappa$ can vary
from generator to generator. It is not difficult to show (see, e.g.,
\cite{w3}) that
\begin{equation}
\kappa = {\rm lim} (Va)
\label{sg}
\end{equation}
where $a$ is the magnitude of the acceleration of the orbits of
$\xi^a$ in the region off of $\cal K$ where they are timelike, $V
\equiv (- \xi^a \xi_a)^{1/2}$ is the ``redshift factor'' of $\xi^a$,
and the limit as one approaches ${\cal K}$ is taken. Equation
(\ref{sg}) motivates the terminology ``surface gravity''. Note that
the surface gravity of a black hole is defined only when it is ``in
equilibrium'', i.e., stationary, so that its event horizon is a
Killing horizon.

In parallel with the two independent ``rigidity theorems'' mentioned
above, there are two independent versions of the zeroth law of black
hole mechanics. The first, due to Carter \cite{c} (see also
\cite{rw2}), states that for any black hole which is static or is
stationary-axisymmetric with the $t - \phi$ orthogonality property,
the surface gravity $\kappa$, must be constant over its event horizon
${\cal H}$. This result is purely geometrical, i.e., it involves no
use of any field equations. The second, due to Bardeen, Carter, and
Hawking \cite{bch} states that if Einstein's equation holds with the
matter stress-energy tensor satisfying the dominant energy condition,
then $\kappa$ must be constant on any Killing horizon. Thus, in the
second version of the zeroth law, the hypothesis that the $t - \phi$
orthogonality property holds is eliminated, but use is made of the
field equations of general relativity.

A {\em bifurcate Killing horizon} is a pair of null surfaces, ${\cal
K}_A$ and ${\cal K}_B$, which intersect on a spacelike 2-surface,
$\cal C$ (called the ``bifurcation surface''), such that ${\cal K}_A$
and ${\cal K}_B$ are each Killing horizons with respect to the same
Killing field $\xi^a$. It follows that $\xi^a$ must vanish on $\cal
C$; conversely, if a Killing field, $\xi^a$, vanishes on a
two-dimensional spacelike surface, ${\cal C}$, then ${\cal C}$ will be
the bifurcation surface of a bifurcate Killing horizon associated with
$\xi^a$ (see \cite{w4} for further discussion). An important
consequence of the zeroth law is that if $\kappa \neq 0$, then in the
``maximally extended'' spacetime representing a stationary black hole,
the event horizon, ${\cal H}$, comprises a branch of a bifurcate
Killing horizon \cite{rw2}. This result is purely
geometrical---involving no use of any field equations. As a
consequence, the study of stationary black holes which satisfy the
zeroth law divides into two cases: ``degenerate'' black holes (for
which, by definition, $\kappa = 0$), and black holes with bifurcate
horizons.

The first law of black hole mechanics is simply an identity relating
the changes in mass, $M$, angular momentum, $J$, and horizon area,
$A$, of a stationary black hole when it is perturbed. To
first order, the variations of these quantities in the vacuum case
always satisfy
\begin{equation}
\delta M = \frac{1}{8 \pi} \kappa \delta A + \Omega \delta J.
\label{bh1}
\end{equation}
In the original derivation of this law \cite{bch}, it was required
that the perturbation be stationary. Furthermore, the original
derivation made use of the detailed form of Einstein's
equation. Subsequently, the derivation has been generalized to hold
for non-stationary perturbations \cite{sw}, \cite{iw}, provided that
the change in area is evaluated at the bifurcation surface, ${\cal
C}$, of the unperturbed black hole. More significantly, it has been
shown \cite{iw} that the validity of this law depends only on very
general properties of the field equations. Specifically, a version of
this law holds for any field equations derived from a diffeomorphism
covariant Lagrangian, $L$. Such a Lagrangian can always be written in
the form
\begin{equation}
L = L \left( g_{ab}; R_{abcd}, \nabla_a R_{bcde},
...;\psi, \nabla_a \psi, ...\right)
\label{lag}
\end{equation}
where $\nabla_a$ denotes the derivative operator associated with
$g_{ab}$, $R_{abcd}$ denotes the Riemann curvature tensor of $g_{ab}$,
and $\psi$ denotes the collection of all matter fields of the theory
(with indices suppressed). An arbitrary (but finite) number of
derivatives of $R_{abcd}$ and $\psi$ are permitted to appear in $L$.
In this more general context, the first law of black hole mechanics is
seen to be a direct consequence of an identity holding for the
variation of the Noether current. The general form of the first law
takes the form
\begin{equation}
\delta M = \frac{\kappa}{2 \pi} \delta S_{\rm bh} + \Omega \delta J + ...,
\label{first}
\end{equation}
where the ``...'' denote possible additional contributions from long
range matter fields, and where
\begin{equation}
S_{\rm bh} \equiv -2 \pi \int_{\cal C} \frac{\delta L}{\delta
R_{abcd}} n_{ab} n_{cd}
\label{Sbh} .
\end{equation}
Here $n_{ab}$ is the binormal to the bifurcation surface $\cal C$
(normalized so that $n_{ab} n^{ab} = -2$), and the functional
derivative is taken by formally viewing the Riemann tensor as a field
which is independent of the metric in eq.(\ref{lag}). For the case of
vacuum general relativity, where $L = R \sqrt{-g}$, a simple
calculation yields
\begin{equation}
S_{\rm bh} = A/4 
\label{Sbh2}
\end{equation}
and eq.(\ref{first}) reduces to eq.(\ref{bh1}).

As already mentioned at the beginning of this section, the black hole
analog of the second law of thermodynamics is the area theorem
\cite{h}. This theorem states that if Einstein's equation holds with
matter satisfying the null energy condition (i.e., $T_{ab} k^a k^b
\geq 0$ for all null $k^a$), then the surface area, $A$, of the event
horizon of a black hole can never decrease with time. In the context
of more general theories of gravity, the nondecrease of $S_{\rm bh}$
also has been shown to hold in a class of higher derivative gravity
theories, where the Lagrangian is a polynomial in the scalar curvature
\cite{jkm2}, but, unlike the zeroth and first laws, no general
argument for the validity of the second law of black hole mechanics is
known. However, there are some hints that the nondecrease of $S_{\rm
bh}$ may hold in a very general class of theories of gravity with
positive energy properties \cite{w2}.

Taken together, the zeroth, first, and second laws\footnote{It should
be noted that I have made no mention of the third law of
thermodynamics, i.e., the ``Planck-Nernst theorem'', which states that
$S \rightarrow 0$ (or a ``universal constant'') as $T \rightarrow
0$. The analog of this law fails in black hole mechanics, since there
exist ``extremal'' black holes of finite $A$ which have $\kappa =
0$. However, I believe that the the ``Planck-Nernst theorem'' should
not be viewed as a fundamental law of thermodynamics but rather as a
property of the density of states near the ground state in the
thermodynamic limit, which is valid for commonly studied
materials. Indeed, examples can be given of ordinary quantum systems
that violate the ``Planck-Nernst theorem'' in a manner very similar to
the violations of the analog of this law that occur for black holes
\cite{w5}.} of black hole mechanics in general relativity are
remarkable mathematical analogs of the corresponding laws in ordinary
thermodynamics. It is true that the nature of the proofs of the laws
of black hole mechanics in classical general relativity is strikingly
different from the nature of the arguments normally advanced for the
validity of the ordinary laws of thermodynamics. Nevertheless, as
discussed above, the validity of the laws of black hole mechanics
appears to rest upon general features of the theory (such as general
covariance) rather than the detailed form of Einstein's equation, in a
manner similar to the way the validity of the ordinary laws of
thermodynamics depends only on very general features of classical and
quantum dynamics.

In comparing the laws of black hole mechanics in classical general
relativity with the laws of thermodynamics, the role of energy, $E$,
is played by the mass, $M$, of the black hole; the role of
temperature, $T$, is played by a constant times the surface gravity,
$\kappa$, of the black hole; and the role of entropy, $S$, is played
by a constant times the area, $A$, of the black hole.  The fact that
$E$ and $M$ represent the same physical quantity provides a strong
hint that the mathematical analogy between the laws of black hole
mechanics and the laws of thermodynamics might be of physical
significance. However, in classical general relativity, the physical
temperature of a black hole is absolute zero, so there can be no
physical relationship between $T$ and $\kappa$. Consequently, it also
would be inconsistent to assume a physical relationship between $S$
and $A$. As we shall now see, this situation changes dramatically when
quantum effects are taken into account.

\section{Quantum black hole thermodynamics}

The physical temperature of a black hole is not absolute zero. As a
result of quantum particle creation effects \cite{h2}, a black hole
radiates to infinity all species of particles with a perfect black
body spectrum, at temperature (in units with $G=c=\hbar=k=1$)
\begin{equation}
T = \frac{\kappa}{2 \pi} .
\label{T}
\end{equation}
Thus, $\kappa/2 \pi$ truly is the {\em physical} temperature of a
black hole, not merely a quantity playing a role mathematically
analogous to temperature in the laws of black hole mechanics. 

In fact, there are two logically independent results which give rise
to the formula (\ref{T}). Although these results are mathematically
very closely related, it is important to distinguish clearly between
them. The first result is the original thermal particle creation
effect discovered by Hawking \cite{h2}. In its most general form, this
result may be stated as follows (see \cite{w4} for further
discussion): Consider a a classical spacetime $(M, g_{ab})$ describing
a black hole formed by gravitational collapse, such that the black
hole ``settles down'' to a stationary final state. By the zeroth law
of black hole mechanics, the surface gravity, $\kappa$, of the black
hole final state will be constant over its event horizon. Consider a
quantum field propagating in this background spacetime, which is
initially in any (non-singular) state. Then, at asymptotically late
times, particles of this field will be radiated to infinity as though
the black hole were a perfect black body\footnote{If the black hole is
rotating, the the spectrum seen by an oberver at infinity corresponds
to what would emerge from a ``rotating black body''.} at the Hawking
temperature, eq. (\ref{T}). It should be noted that this result relies
only on the analysis of quantum fields in the region exterior to the
black hole, and it does not make use of any gravitational field
equations.

The second result is the Unruh effect \cite{u} and its generalization
to curved spacetime. In its most general form, this result may be
stated as follows (see \cite{kw}, \cite{w4} for further discussion):
Consider a a classical spacetime $(M, g_{ab})$ that contains a
bifurcate Killing horizon, ${\cal K} = {\cal K}_A \cup {\cal K}_B$,
i.e., there is a one-parameter group of isometries whose associated
Killing field, $\xi^a$, is normal to ${\cal K}$. Consider a free
quantum field on this spacetime. Then there exists at most one
globally nonsingular state of the field which is invariant under the
isometries. Furthermore, in the ``wedges'' of the spacetime where the
isometries have timelike orbits, this state (if it exists) is a KMS
(i.e., thermal equilibrium) state at temperature (\ref{T}) with
respect to the isometries.

Note that in Minkowski spacetime, any one-parameter group of Lorentz
boosts has an associated bifurcate Killing horizon, comprised by two
intersecting null planes. The unique, globally nonsingular state which
is invariant under these isometries is simply the usual (``inertial'')
vacuum state, $|0>$.  In the ``right and left wedges'' of Minkowski
spacetime defined by the Killing horizon, the orbits of the Lorentz
boost isometries are timelike, and, indeed, these orbits correspond to
worldlines of uniformly accelerating observers. If we normalize the
boost Killing field, $b^a$, so that Killing time equals proper time on
an orbit with acceleration $a$, then the surface gravity of the
Killing horizon is $\kappa = a$. An observer following this orbit
would naturally use $b^a$ to define a notion of
``time translation symmetry''. Consequently, when the field is in the
inertial vacuum state, a uniformly accelerating observer would
describe the field as being in a thermal equilibrium state at
temperature
\begin{equation}
T = \frac{a}{2 \pi} 
\label{Tu}
\end{equation}
as originally found by Unruh \cite{u}.

Although there is a close mathematical relationship between the two
results described above, it should be emphasized these results refer
to {\em different} states of the quantum field. In the Hawking effect,
the asymptotic final state of the quantum field is a state in which
the modes of the quantum field that appear to a distant observer to
have propagated from the black hole region of the spacetime are
thermally populated at temperature (\ref{T}), but the modes which
appear to have propagated in from infinity are unpopulated. This state
(usually referred to as the ``Unruh vacuum'') would be singular on the
white hole horizon in the analytically continued spacetime containing
a bifurcate Killing horizon. On the other hand, in the Unruh effect
and its generalization to curved spacetimes, the state in question
(usually referred to as the ``Hartle-Hawking vacuum'') is globally
nonsingular, and {\em all} modes of the quantum field in the ``left
and right wedges'' are thermally populated.\footnote{The state in
which none of the modes in the region exterior to the black hole are
populated is usually referred to as the ``Boulware vacuum''. The
Boulware vacuum is singular on both the black hole and white hole
horizons.}

It also should be emphasized that in the Hawking effect, the
temperature (\ref{T}) represents the temperature as measured by an
observer near infinity. For any observer following an orbit of the
Killing field, $\xi^a$, normal to the horizon, the locally measured
temperature of the modes which appear to have propagated from the
direction of the black hole is given by
\begin{equation}
T = \frac{\kappa}{2 \pi V} \: ,
\label{ta}
\end{equation}
where $V = (-\xi^a \xi_a)^{1/2}$. In other words, the locally measured
temperature of the Hawking radiation follows the Tolman law. Now, as
one approaches the horizon of the black hole, the modes which appear
to have propagated from the black hole dominate over the modes which
appear to have propagated in from infinity. Taking eq.(\ref{sg}) into
account, we see that $T \rightarrow a/2 \pi$ as the black hole
horizon, $\cal H$, is approached, i.e., in this limit eq.(\ref{ta})
corresponds to the flat spacetime Unruh effect.

Equation (\ref{ta}) shows that when quantum effects are taken into
account, a black hole is surrounded by a ``thermal atmosphere'' whose
local temperature as measured by observers following orbits of $\xi^a$
becomes divergent as one approaches the horizon. As we shall see
explicitly below, this thermal atmosphere produces important physical
effects on quasi-stationary bodies near the black hole. On the other
hand, for a macroscopic black hole, observers who freely fall into the
black hole would not notice any important quantum effects as they
approach and cross the horizon.

The fact that $\kappa / 2 \pi$ truly represents the physical
temperature of a black hole provides extremely strong evidence that
the laws of black hole mechanics are not merely mathematical analogs
of the laws of thermodynamics, but rather that they in fact {\em are}
the ordinary laws of themodynamics applied to black holes. If so, then
$A/4$ must represent the physical entropy of a black hole in general
relativity. What is the evidence that this is the case?

Although quantum effects on matter fields outside of a black hole were
fully taken into account in the derivation of the Hawking effect,
quantum effects of the gravitational field itself were not, i.e., the
Hawking effect is derived in the context of semiclassical gravity,
where the effects of gravitation are still represented by a classical
spacetime. As discussed further below, a proper accounting of the
quantum degrees of freedom of the gravitational field itself
undoubtedly would have to be done in order to understand the origin of
the entropy of a black hole. Nevertheless, as I shall now describe,
even in the context of semiclassical gravity, I believe that there are
compelling arguments that $A/4$ must represent the physical entropy of
a black hole.

Even within the semi-classical approximation, conservation of energy
requires that an isolated black hole must lose mass in order to
compensate for the energy radiated to infinity by the particle
creation process. If one equates the rate of mass loss of the black
hole to the energy flux at infinity due to particle creation, one
arrives at the startling conclusion that an isolated black hole will
radiate away all of its mass within a finite time. During this process of
black hole ``evaporation'', $A$ will decrease, in violation of the
second law of black hole mechanics. Such an area decrease can occur
because the expected stress-energy tensor of quantum matter does not
satisfy the null energy condition---even for matter for which this
condition holds classically---in violation of a key hypothesis of the
area theorem. Thus, it is clear that the second law of black hole
mechanics must fail when quantum effects are taken into account.

On the other hand, there is a serious difficulty with the ordinary
second law of thermodynamics when black holes are present: One can
simply take some ordinary matter and drop it into a black hole, where,
classically at least, it will disappear into a spacetime
singularity. In this latter process, one loses the entropy initially
present in the matter, but no compensating gain of ordinary entropy
occurs, so the total entropy, $S$, of matter in the universe decreases.

Note, however, that in the black hole evaporation process, although
$A$ decreases, there is significant amount of ordinary entropy
generated outside the black hole due to particle creation. Similarly,
when ordinary matter (with positive energy) is dropped into a black
hole, although $S$ decreases, by the first law of black hole
mechanics, there will necessarily be an increase in $A$.  These
considerations motivated the following proposal \cite{b1}, \cite{b2}.
Perhaps in any process, the total generalized entropy, $S'$, never
decreases
\begin{equation}
\Delta S' \geq 0 ,
\label{gsl}
\end{equation}
where $S'$ is defined by 
\begin{equation}
S' \equiv S + A/4 .
\label{S'}
\end{equation}

It is not difficult to see that the generalized second law holds for
an isolated black hole radiating into otherwise empty space. However,
it is not immediately obvious that it holds if one carefully lowers a
box containing matter with entropy $S$ and energy $E$ toward a black
hole. Classically, if one lowers the box sufficiently close to the
horizon before dropping it in, one can make the increase in $A$ as
small as one likes while still getting rid of all of the entropy, $S$,
originally in the box. However, it is here that the quantum ``thermal
atmosphere'' surrounding the black hole comes into play. The
temperature gradient in the thermal atmosphere (see
eq.(\ref{ta})) implies that there is a pressure gradient and,
consequently, a buoyancy force on the box. As a result of this
buoyancy force, the optimal place to drop the box into the black hole
is no longer the horizon but rather the ``floating point'' of the box,
where its weight is equal to the weight of the displaced thermal
atmosphere. The minimum area increase given to the black hole in the
process is no longer zero, but rather it turns out to be an amount
just sufficient to prevent any violation of the generalized second law
from occurring \cite{uw}. A number of other analyses \cite{tz},
\cite{fp}, \cite{s} also have given strong support to validity of the
generalized second law.

The generalized entropy (\ref{S'}) and the generalized second law
(\ref{gsl}) have obvious interpretations: Presumably, for a system
containing a black hole, $S'$ is nothing more than the ``true total
entropy'' of the complete system, and (\ref{gsl}) is then nothing more
than the ``ordinary second law'' for this system. If so, then $A/4$
truly is the physical entropy of a black hole.  

I believe that the above semi-classical considerations make a
compelling case for the merger of the laws of black hole mechanics
with the laws of thermodynamics. However, if one is to obtain a
deeper understanding of why $A/4$ represents the entropy of a black
hole in general relativity, it clearly will be necessary to go beyond
semi-classical considerations and attain an understanding of the
quantum dynamical degrees of freedom of a black hole. Thus, one would
like to calculate the entropy of a black hole directly from a quantum
theory of gravity. There have been many attempts to do so, most of
which fall within the following categories: (i) Calculations that are
mathematically equivalent to the classical calculation described in
the previous section. (ii) Calculations that ascribe a preferred local
significance to the horizon. (iii) State counting calculations of
configurations that can be associated with black holes.

The most prominent of the calculations in category (i) is the
derivation of black hole entropy in Euclidean quantum gravity, originally
given by Gibbons and Hawking \cite{gh}. Here one starts with a formal,
functional integral expression for the partition function in Euclidean
quantum gravity and evaluates it for a black hole in the ``zero loop''
(i.e, classical) approximation. As shown in \cite{w6}, the
mathematical steps in this procedure are in direct correspondence with
the purely classical determination of the entropy from the form of the
first law of black hole mechanics. Thus, although this derivation
gives some intriguing glimpses into possible deep relationships
between black hole thermodynamics and Euclidean quantum gravity, the
Euclidean derivation does not appear to provide any more insight than
the classical derivation into accounting for the quantum degrees of
freedom that are responsible for black hole entropy. Similar remarks
apply to a number of other entropy calculations that also can be shown
to be equivalent to the classical derivation (see \cite{iw2}).

Within category (ii), a key approach has been to attribute the entropy
of the black hole to ``entanglement entropy'' resulting from quantum
field correlations between the exterior and interior of the black hole
(see, in particular, \cite{bkls}). As a result of these correlations,
the state of the field when restricted to the exterior of the black
hole is mixed, and its von Neumann entropy, $- {\rm tr} [\hat{\rho}
\ln \hat{\rho}]$, would diverge in the absence of a short distance
cutoff. If one now inserts a short distance cutoff of the order of the
Planck scale, one obtains a von Neumann entropy of the order of the
horizon area, $A$. A closely related idea is to attribute the entropy
of the black hole to the ordinary entropy of its thermal atmosphere
(see, in particular, \cite{th}). Since $T$ diverges near the horizon
in the manner specified by eq.(\ref{ta}), the entropy of the thermal
atmosphere diverges, but if one puts in a Planck scale cutoff, one
gets an entropy of order $A$. Indeed, this calculation is really the
same as the entanglement entropy calculation, since the state of a
quantum field outside of the black hole at late times is thermal, so
its von Neumann entropy is equal to its thermodynamic entropy.

These and other approaches in category (ii) provide a natural way of
accounting for why the entropy of a black hole is proportional to its
surface area, although the constant of proportionality typically
depends upon a cutoff or other free parameter and is not
calculable. However, it is far from clear why the black hole horizon
should be singled out for such special treatment of the quantum
degrees of freedom in its vicinity, since, for example, similar
quantum field correlations will exist across any other null
surface. Indeed, as discussed further at the end of the next section,
it is particularly puzzling why the local degrees of freedom
associated with the horizon should be singled out since, as already
noted above, the black hole horizon at a given time is defined in
terms of the entire future history of the spacetime and thus has no
distinguished local significance. Finally, for approaches in category
(ii) that do not make use of the gravitational field equations---such
as the ones described above---it is difficult to see how one would
obtain a black hole entropy proportional to eq.(\ref{Sbh}) (rather
than proportional to $A$) in a more general theory of gravity.

By far, the most successful calculations of black hole entropy to date
are ones in category (iii) that obtain the entropy of certain extremal
and nearly extremal black holes in string theory. It is believed that
at ``low energies'', string theory should reduce to a 10-dimensional
supergravity theory. If one treats this supergravity theory as a
classical theory involving a spacetime metric, $g_{ab}$, and other
classical fields, one can find solutions describing black holes. On
the other hand, one also can consider a ``weak coupling'' limit of
string theory, wherein the states are treated perturbatively about a
background, flat spacetime.  In the weak coupling limit, there is no
literal notion of a black hole, just as there is no notion of a black
hole in linearized general relativity. Nevertheless, certain weak
coupling states can be identified with certain black hole solutions of
the low energy limit of the theory by a correspondence of their energy
and charges. (Here, it is necessary to introduce ``D-branes'' into
string perturbation theory in order to obtain weak coupling states
with the desired charges.) Now, the weak coupling states are, in
essence, ordinary quantum dynamical degrees of freedom in a flat
background spacetime, so their entropy can be computed by the usual
methods of flat spacetime statistical physics.  Remarkably, for
certain classes of extremal and nearly extremal black holes, the
ordinary entropy of the weak coupling states agrees exactly with the
expression for $A/4$ for the corresponding classical black hole
states; see \cite{ho} for a review of these results.

Since the formula for entropy has a nontrivial functional dependence
on energy and charges, it is hard to imagine that this agreement
between the ordinary entropy of the weak coupling states and black
hole entropy could be the result of a random coincidence. Furthermore,
for low energy scattering, the absorption/emission coefficients
(``gray body factors'') of the corresponding weak coupling states and
black holes also agree \cite{ms}. This suggests that there may be a
close physical association between the weak coupling states and black
holes, and that the dynamical degrees of freedom of the weak coupling
states are likely to at least be closely related to the dynamical
degrees of freedom responsible for black hole entropy. However, it
seems hard to imagine that the weak coupling states could be giving an
accurate picture of the local physics occurring near (and within) the
region classically described as a black hole. Thus, it seems likely
that in order to attain additional new conceptual insights into the
nature of black hole entropy in string theory, further significant
progress will have to be made toward obtaining a proper local
description of strong gravitational field phenomena.

\section{Some unresolved issues and puzzles}

I believe that the results described in the previous two sections
provide a remarkably compelling case that black holes are localized
thermal equilibrium states of the quantum gravitational
field. Although none of the above results on black hole thermodynamics
have been subject to any experimental or observational tests, the
theoretical foundation of black hole thermodynamics is sufficiently
firm that I feel that it provides a solid basis for further research
and speculation on the nature of quantum gravitational
phenomena. Indeed, it is my hope that black hole thermodynamics will
provide us with some of the additional key insights that we will need
in order to gain a deeper understanding of quantum gravitational
phenomena. In this section, I will raise and discuss four major,
unresolved issues in quantum gravitational physics that black hole
thermodynamics may help shed light upon.

\medskip

\noindent
{\bf I.} {\it What is the nature of singularities in quantum gravity?}

The singularity theorems of classical general relativity assert that
in a wide variety of circumstances, singularities must occur in the
sense that spacetime cannot be geodesically complete. However,
classical general relativity should break down prior to the formation
of a singularity. One possibility is that in quantum gravity, these
singularities will be ``smoothed over''. However, it also is possible
that at least some aspects of the singularities of classical general
relativity are true features of nature, and will remain present in
quantum gravitational physics.

Black hole thermodynamics provides a strong argument that the
singularity inside of a black hole in classical general relativity
will remain present in at least some form in quantum gravity. In
classical general relativity, the matter responsible for the formation
of the black hole propagates into a singularity in the deep interior
of the black hole. Suppose that the matter which forms the black hole
possesses quantum correlations with matter that remains far outside of
the black hole. Then it is hard to imagine how these correlations
could be restored during the process of black hole evaporation;
indeed, if anything, the Hawking process should itself create
additional correlations between the exterior and interior of the black
hole as it evaporates (see \cite{w4} for further discussion). However,
if these correlations are not restored, then by the time that the
black hole has evaporated completely, an initial pure state will have
evolved to a mixed state, i.e., ``information'' will have been
lost. In the semiclassical picture, such information loss does occur
and is ascribable to the propagation of the quantum correlations into
the singularity within the black hole. If pure states continue to
evolve to mixed states in a fully quantum treatment of the
gravitational field, then at least the aspect of the classical
singularity as a place where ``information can get lost'' must remain
present in quantum gravity. This issue is frequently referred to as
the ``black hole information paradox'', and its resolution would tell
us a great deal about the existence and nature of singularities in
quantum gravity.

\medskip

\noindent
{\bf II}. {\it Is there a relationship between singularities and the
second law?}

The usual arguments for the validity of the second law of
thermodynamics rest upon having very ``special'' (i.e., low entropy)
initial conditions. Such special initial conditions in systems that we
presently observe trace back to even more special conditions at the
(classically singular) big bang origin of the universe. Thus, the
validity of the second law of thermodynamics appears to be intimately
related to the nature of the initial singularity \cite{p}. On the
other hand, the arguments leading to the area increase theorem for
black holes in classical general relativity would yield an area
decrease theorem if applied to white holes. Thus, the applicability
(or, at least, the relevance) of the second law of black hole
mechanics appears to rest upon the fact that black holes can occur in
nature but white holes do not. This, again, could be viewed as a
statement about the types of singularities that can occur in nature
\cite{p}.  If, as argued here, the laws of black hole mechanics {\em
are} the laws of thermodynamics applied to a system containing a black
hole, then it seems hard to avoid the conclusion that a close
relationship must exist between the second law of thermodynamics and
the nature of what we classically describe as singularities.

\medskip

\noindent
{\bf III}. {\it Are statistical probabilities truly distinct from
quantum probabilities?}

Even in classical physics, probablities come into play in statistical
physics as ensembles representing our ignorance of the exact state of
the system. On the other hand, in quantum physics, probabilities enter
in a much more fundamental way: Even if the state of a system is
known exactly, one can only assign a probability distribution to the
value of observables. In quantum statistical physics, probabilities
enter in both of these ways, and it would seem that these two ways
should be logically distinguishable. However, density matrices have
the odd feature of entering quantum statistical physics in two
mathematically equivalent ways: (i) as an exact description of a
particular (mixed) quantum state, and (ii) as a statistical ensemble of a
collection of pure quantum states. In particular, one may choose to
view a thermal density matrix either as a single, definite (mixed)
state of the quantum system, or as a statistical ensemble of pure
states. In the former case, the probability distribution for the
values of observables would be viewed as entirely quantum in origin,
whereas in the latter case, it would be viewed as partly statistical
and partly quantum in origin; indeed, for certain observables (such as
the energy of the system), the probabilities in the second case would
be viewed as entirely statistical in origin. The Unruh effect puts
this fact into a new light: When a quantum field is in the ordinary
vacuum state, $|0>$, it is in a pure state, so the probability
distribution for any observable would naturally be viewed by an
inertial observer to be entirely quantum in origin. On the other hand,
for an accelerating observer, the field is in a thermal state at
temperature (\ref{Tu}), and the probability distribution for
``energy'' (conjugate to the notion of time translation used by the
accelerating observer) would naturally be viewed as entirely
statistical in origin. Although there are no physical or mathematical
inconsistencies associated with these differing viewpoints, they seem
to suggest that there may be some deep connections between quantum
probabilities and statistical probablities; see \cite{sm} for further
exploration of these ideas.

\medskip

\noindent
{\bf IV}. {\it What is the definition/meaning of entropy in general
relativity?}

The issue of how to assign entropy to the gravitational field has been
raised and discussed in the literature (see, in particular, \cite{p}),
although it seems clear that a fully quantum treatment of the degrees
of freedom of the gravitational field will be essential for this issue
to be resolved. However, as I will emphasize below, even the
definition and meaning of the entropy of ``ordinary matter'' in
general relativity raises serious issues of principle, which have
largely been ignored to date.

First, it should be noted that underlying the usual notion of entropy
for an ``ordinary system'' is the presence of a well defined notion of
``time translations'', which are symmetries of the dynamics. The total
energy of the system is then well defined and conserved. The
definition and meaning of the usual notion of entropy for classical
systems is then predicated on the assumption that generic dynamical
orbits ``sample'' the entire energy shell, spending ``equal times in
equal volumes''; a similar assumption underlies the notion of entropy
for quantum systems (see \cite{w2} for further discussion). Now, an
appropriate notion of ``time translations'' is present when one
considers dynamics on a background spacetime whose metric possesses a
suitable one-parameter group of isometries, and when the Hamiltonian
of the system is invariant under these isometries. However, such a
structure is absent in general relativity, where no background metric
is present.\footnote{Furthermore, it is clear that gross violations of
any sort of ``ergodic behavior'' occur in classical general relativity
on account of the irreversible tendency for gravitational collapse to
produce singularities, from which one cannot then evolve back to
uncollapsed states---although the semiclassical process of black hole
evaporation suggests the possibility that ergodic behavior could be
restored in quantum gravity.} The absence of any ``rigid'' time
translation structure can be viewed as being
responsible for making notions like the ``energy density of the
gravitational field'' ill defined in general relativity. Notions like
the ``entropy density of the gravitational field'' are not likely to
fare any better. It may still be possible to use structures like
asymptotic time translations to define the notion of the total entropy
of an (asymptotically flat) isolated system. (As is well known, total
energy can be defined for such systems.) However, for a closed
universe, it seems highly questionable that any meaningful notion will
exist for the ``total entropy of the universe'' (including
gravitational entropy).

The comments in the previous paragraph refer to serious difficulties
in defining the notions of gravitational entropy and total entropy in
general relativity. However, as I now shall explain, even in the
context of quantum field theory on a background spacetime possessing a
time translation symmetry---so that the ``rigid'' structure needed to
define the usual notion of entropy of matter is present---there are
strong hints from black hole thermodynamics that even our present
understanding of the meaning of the ``ordinary entropy'' of matter is
inadequate.

Consider the ``thermal atmosphere'' of a black hole. As discussed in
Section 3 above, since the locally measured temperature is given by
eq.(\ref{ta}), if we try to compute its ordinary entropy, a new
ultraviolet catastrophe occurs: The entropy is infinite unless we put
in a cutoff on the contribution from short wavelength
modes.\footnote{Since a field has infinitely many degrees of freedom,
it threatens to make an infinite contribution to entropy. The old
ultraviolet catastrophe---which plagued physics at the turn of the
previous century---was resolved by quantum theory, which, in effect,
provides a cutoff on the entropy contribution of modes with energy
greater than $kT$, so that, at any $T$, only finitely many degrees of
freedom are relevant. The new ultraviolet catastrophe arises because,
on account of arbitrarily large redshifts, there now are infinitely
many modes with energy less than $kT$. To cure it, it is necessary to
have an additional cutoff (presumably arising from quantum gravity) on
short wavelength modes.} As already noted in Section 3, if we insert a
cutoff of the order of the Planck scale, then the thermal atmosphere
contributes an entropy of order the area, $A$, of the horizon (in
Planck units). Note that the bulk of the entropy of the thermal
atmosphere is highly localized in a ``skin'' surrounding the horizon,
whose thickness is of order of the Planck length. The presence of this
thermal atmosphere poses the following puzzle:

{\bf Puzzle}: {\it What is the ``physical entropy'' of the thermal
atmosphere?}

One possibility is that the thermal atmosphere should be assigned an
entropy of order the area of the horizon, as indicated above. As
discussed in Section 3, this would then account (in order of
magnitude) for the entropy of black holes. However, this also would
mean that there would be no room left to assign entropy to any
``internal degrees of freedom'' of the black hole, i.e., all of the
entropy of a black hole would be viewed as residing in a Planck scale
skin surrounding the horizon. To examine the implications of this view
in a more graphic manner, consider the collapse of a very massive
spherical shell of matter, say of mass $M = 10^{11} M_{\odot}$. Then,
as the shell crosses its Schwarzschild radius, $R \sim 3 \times
10^{11} {\rm km}$, the spacetime curvature outside of the shell is
much smaller than that at the surface of the Earth, and it will take
more than another week before the shell collapses to a singularity. An
unsophisticated observer riding on the shell would have no idea that
any doom awaits him, and he would notice nothing of any significance
occurring as the Schwarzschild radius is crossed. Nevertheless, within
a time of order the Planck time after crossing of the Schwarzschild
radius, the ``skin'' of thermal atmosphere surrounding the newly
formed black hole will come to equilibrium with respect to the notion
of time translation symmetry for the static Schwarzschild
exterior. Thus, if entropy is to be assigned to the thermal atmosphere
as above, then the degrees of freedom of the thermal
atmosphere---which previously were viewed as irrelevant vacuum
fluctuations making no contribution to entropy---suddenly become
``activated'' by the passage of the shell for the purpose of counting
their entropy. A momentous change in the entropy of matter in the
universe has occurred, and all of this entropy increase is localized
near the Schwarzschild radius of the shell, but the observer riding
the shell sees nothing.\footnote{Similarly, if the entropy of the
thermal atmosphere is to be taken seriously, then it would seem that
during a period of uniform acceleration, an observer in Minkowski
spacetime should assign an infinite entropy (since the horizon area is
infinite) to a Planck sized neighborhood of a pair of intersecting
null planes lying at a distance $c^2/a$ from him. Observers near these
null planes presumably would be quite surprised by the assignment of a
huge entropy density to an ordinary, empty region of Minkowski
spacetime.}

Another possibility is that the infinite (prior to the imposition of a
cutoff) entropy of the thermal atmosphere is simply another infinity
of quantum field theory that needs to be properly ``renormalized'';
when a proper renormalization has been done, the thermal atmosphere
will make a negligible contribution to the total entropy. This view
would leave room to attribute black hole entropy to ``internal degrees
of freedom of the black hole'', and would avoid the difficulties
indicated in the previous paragraph. However, it raises serious new
difficulties of its own. Consider a black hole enclosed in a
reflecting cavity which has come to equilibrium with its Hawking
radiation. Surely, far from the black hole, the entropy of the thermal
radiation in the cavity should not be ``renormalized away''. But this
radiation is part of the thermal atmosphere of the black hole. Thus,
one would have to postulate that at some distance from the black hole,
the renormalization effects begin to become important. In order to
avoid the difficulties of the previous paragraph, this would have to
occur at a distance much larger than the Planck length. But, then,
what happens to the entropy in a box of ordinary thermal matter as it
is slowly lowered toward the black hole. By the time it reaches its
``floating point'', its contents are indistinguishable from the
thermal atmosphere. Thus, if the floating point is close enough to the
black hole for the renormalization to have occurred, the entropy in
the box must have disappeared, despite the fact that an observer
inside the box still sees it filled with thermal radiation.
Furthermore, if one lowers (or, more accurately, pushes) an empty box
to the same distance from the black hole, it will have an entropy less
than the box filled with radiation. Therefore, the empty box would
have to be assigned a negative entropy.

\bigskip

I believe that the above puzzle suggests that we presently lack the
proper conceptual framework with which to think about entropy in the
context of general relativity. In any case, it is my belief that the
resolution of the above issues will occupy researchers well into the
next century, if not well into the next millenium.

This research was supported in part by NSF grant PHY 95-14726 to the
University of Chicago.

\end{document}